\documentclass[12pt]{iopart}

\usepackage{amsmath, amssymb}  
\usepackage[british]{babel}
\usepackage[utf8]{inputenc}
\usepackage{graphicx}
\usepackage[per-mode=symbol]{siunitx}
\usepackage{hyperref}

\newcommand{\erfc}[1]{\mathrm{erfc}\left(#1\right)}
\newcommand{\G}{\mathcal G}
\newcommand{\ldiff}{l_\text{diff}}
\newcommand{\lamdiff}{\lambda_\text{diff}}

\renewcommand{\vec}[1]{\mathbf{#1}}

\begin{document}

\title[Photon BEC at Crossover]{Photon BEC with Thermo-Optic Interaction\\ at Dimensional Crossover}

\author{Enrico Stein}
\ead{estein@rhrk.uni-kl.de}
\address{Department of Physics and Research Center OPTIMAS, Technische Universität Kaiserslautern, Erwin-Schrödinger Straße 46, 67663 Kaiserslautern, Germany}

\author{Axel Pelster}
\ead{axel.pelster@physik.uni-kl.de}
\address{Department of Physics and Research Center OPTIMAS, Technische Universität Kaiserslautern, Erwin-Schrödinger Straße 46, 67663 Kaiserslautern, Germany}

\vspace{10pt}
\begin{indented}
\item[]\today
\end{indented}

\begin{abstract}
Since the advent of experiments with photon Bose-Einstein condensates in dye-filled microcavities in 2010, many investigations have focused upon the emerging effective photon-photon interaction. Despite its smallness, it can be identified to stem from two physically distinct mechanisms. On the one hand, a Kerr nonlinearity of the dye medium yields a photon-photon contact interaction. On the other hand, a heating of the dye medium leads to an additional thermo-optic interaction, which is both delayed and non-local. The latter turns out to represent the leading contribution to the effective interaction for the current 2D experiments.\\
Here we analyse theoretically how the effective photon-photon interaction increases when the system dimension is reduced from 2D to 1D. To this end, we consider an anisotropic harmonic trapping potential and determine via a variational approach how the properties of the photon Bose-Einstein condensate in general, and both aforementioned interaction mechanisms in particular, change with increasing anisotropy. We find that the thermo-optic interaction strength increases at first linearly with the trap aspect ratio and lateron saturates at a certain value of the trap aspect ratio. Furthermore, in the strong 1D limit the roles of both interactions get reversed as the thermo-optic interaction remains saturated and the contact Kerr interaction becomes the leading interaction mechanism. Finally, we discuss how the predicted effects can be measured experimentally.
\end{abstract}

%
\vspace{2pc}
%
\noindent{\submitto{\NJP}}
%
\maketitle

\section{Introduction}
\begin{figure}[t]
	\centering
	\includegraphics[width=.5\linewidth]{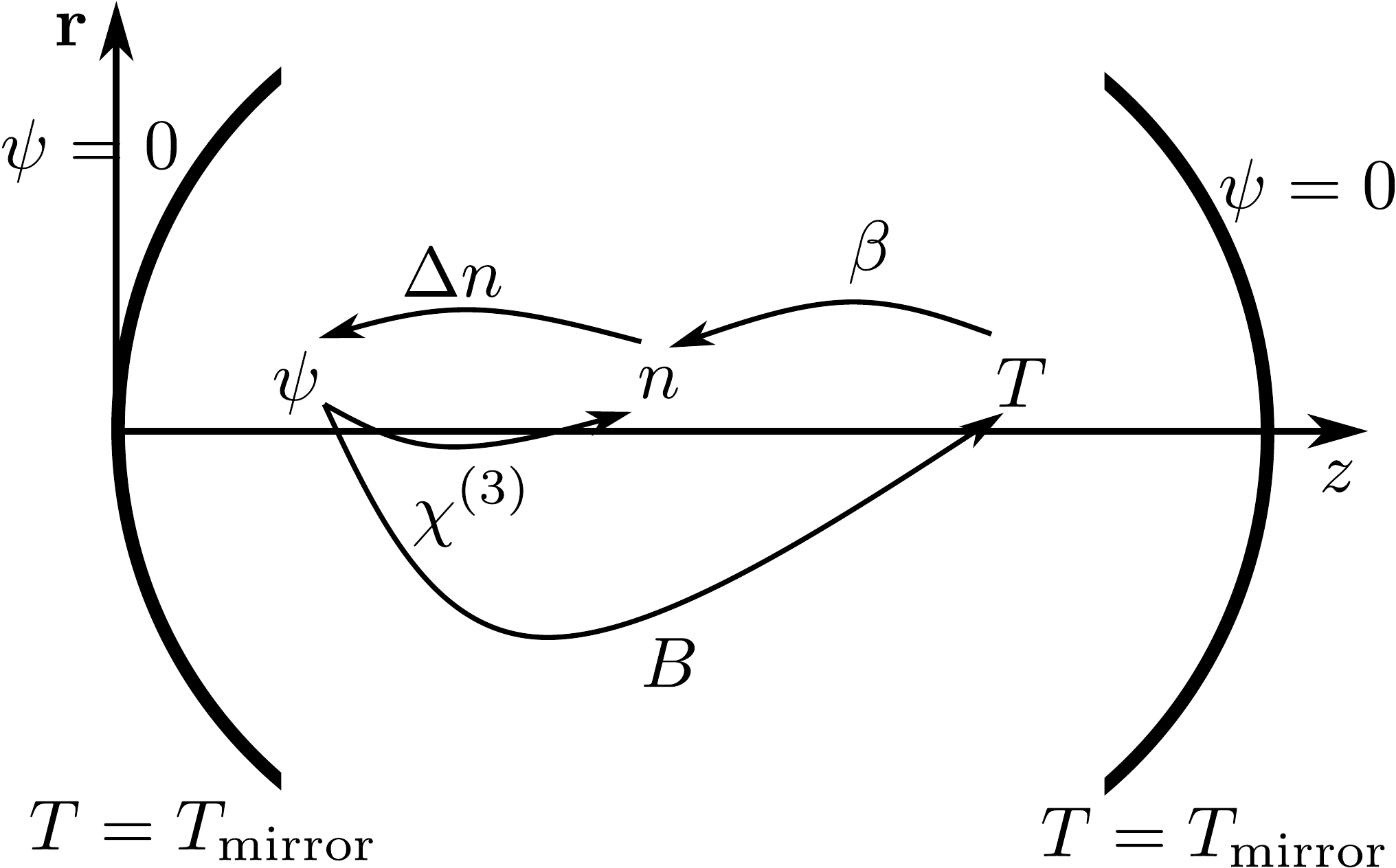}
	\caption{Sketch of the interaction mechanisms mediated by the dye solution. The change $\Delta n$ of the refractive index $n$ of the dye solution stems from
	both a Kerr nonlinearity $\chi^{(3)}$ and thermo-optics described by the coefficient $\beta$. The picture is taken from reference \cite{Stein2019}.}
	\label{Fig:interaction}
\end{figure}

Ultracold atomic quantum systems in dimensions lower than three bear interesting physics \cite{RevModPhys.80.885,Giamarchi}. In 2D an interacting Bose gas can undergo a crossover from a Bose-Einstein condensate to a Berezinskii–Kosterlitz–Thouless (BKT) phase \cite{Hadzibabic2006,PhysRevLett.114.255302, Christodoulou2021}, where vortex and anti-vortex pairs are produced and can move through the gas. In one-dimensional systems large phase fluctuations are detected \cite{Dettmer2001}, which induce an algebraic decay of the correlation function, in contrast to an exponential decay in higher dimensions. Moreover, it is also known that the effective interaction strength increases by reducing the dimension of the system \cite{Cazalilla2006}. For systems of bosonic atoms the dimensional crossover has already been investigated broadly from 3D to 2D \cite{PhysRevA.96.043623} and even down to 1D \cite{PhysRevLett.113.215301,PhysRevA.95.043610}. The question of the effective system dimension can be reduced to a discussion of the relevant length scales \cite{Gorlitz2001}. Provided that the healing length of a three-dimensional condensate axially symmetrical trap is larger than the axial width, the system is effectively two-dimensional. In case that the in-plane radius is smaller than the healing length, the system is quasi-1D.\\
For photon Bose-Einstein condensates (phBEC) \cite{Klaers2010}, however, such a dimensional cross\-over has so far not been realised. As these kinds of experiments are conducted in a microcavity, they turn out to be already two-dimensional. It is expected that the crossover to 1D can be achieved experimentally by writing an anisotropic harmonic confining potential directly on the mirror \cite{Maruo97, Deubel2004, Hohmann2015}. This should yield a simple control of the trap anisotropy, which then allows to freeze out the higher dimension as has already been shown in the theoretical study \cite{Stein2021}. Thus, such photonic systems constitute a useful platform to investigate the crossover from higher to lower dimensions. \\
In the corresponding experimental set-up, photons are trapped in a dye-filled cavity and, due to the contact with the dye, the photon gas is allowed to thermalise \cite{nat-phys} and finally to Bose-Einstein
condense \cite{Klaers2010}. Moreover, the dye solution leads also to an effective photon-photon interaction via two mechanisms, as is depicted in figure~\ref{Fig:interaction}.
One is the Kerr effect, which is due to a nonlinearity $\chi^{(3)}$ of the solvent molecules, where a change of the refractive index $\Delta n\propto|\psi|^2$ leads to an effective contact interaction. The second mechanism for the effective photon-photon interaction is the thermo-optic effect. Here the dye solution heats up due to non-perfect absorption-reemission cycles and some photons are converted into excitations of the dye solution, leading to its net heating. This changes the refractive index of the dye solution according to the thermo-optic coefficient $\beta$ and, thus, contributes to the effective photon-photon interaction.\\
It turns out that the thermo-optic interaction is the leading contribution in the current 2D experiments. However, the total interaction strength is still quite small as the dimensionless interaction strength amounts to about $\tilde g=mg/\hbar^2\sim 10^{-4}$ \cite{Klaers2010,ApplPhysB}. Therefore, effects of stronger interaction like superfluidity are not yet observable and even the thermodynamics turns out to be not affected by the interaction \cite{Damm2016}. This finding motivated our previous study \cite{Stein2021}, where we investigated as a first step the dimensional crossover of a non-interacting photon BEC from 2D to 1D by determining its thermodynamic properties and by extracting from them the effective system dimension for a given temperature and trap aspect ratio. In a second step, it is now crucial to search for mechanisms to increase the effective photon-photon interaction. In this respect we already found in the former theoretical study \cite{Stein2019} the intriguing result that the strength of the thermo-optic interaction increases quadratically with the lateral extension of the cavity mirrors. However, as this would be quite laborious to achieve experimentally, we explore here an alternative mechanism, which relies on increasing the effective photon-photon interaction strength by reducing the system dimension from 2D to 1D. As it is already known that this increases effectively the contact interaction \cite{PhysRevA.65.043614}, our main focus lies hereby on the question how the dimensional crossover modifies the thermo-optic interaction.\\
To this end, we start by introducing in section \ref{sec:general} a coupled system of mean-field equations describing the steady state of both the phBEC ground state and the temperature, which is produced by the phBEC and which conversely affects the photon-photon interaction. Instead of straight-forwardly solving this coupled system of equations by numerical means, we construct an approximate solution within a semi-analytic procedure as follows. At first, we eliminate the temperature degrees of freedom by using the corresponding Green's function and determine with this the resulting energy functional for the condensate. As the profile of the photon condensate wave function is a Gaussian in the non-interacting case, it is reasonable to assume that this profile remains to be valid also in the mutual presence of both Kerr and thermo-optic interaction. Therefore, within a variational approach, we minimise the condensate energy function with respect to the widths of the used Gaussian trial function in section \ref{sec:minimisation}. Solving the corresponding self-consistency conditions for these widths, it turns out that the dimensional crossover can physically be divided into three different regimes. The first one corresponds to small trap aspect ratios $\lambda$ and shows, as expected, an increase of the thermo-optic interaction strength. In the second regime for intermediate $\lambda$, the thermo-optic interaction turns out to saturate, as here the condensate width in the squeezed direction is smaller than the characteristic length scale of the temperature diffusion. Finally, in the third regime for large $\lambda$ the contact Kerr interaction turns out to take over the leading role in the effective photon-photon interaction. At the end, we discuss that the respective strengths of Kerr and thermo-optic interaction can not only be extracted from the condensate widths but also from analysing the energy in the quasi 1D regime.

\section{General Equations} \label{sec:general}
Our starting point for describing the photon BEC ground state is the mean-field theory worked out in reference~\cite{Stein2019}, see figure~\ref{Fig:interaction}. There we used a set of two coupled equations in order to describe both the photon BEC wave function in the microcavity and the heat diffusion in the dye solution inducing the thermo optics. However, for the current purpose, we consider two modifications of this mean-field theory. On the one hand, we neglect the imaginary part of the equation for the condensate wave function, as this simply determines the photon number $N$. On the other hand, we also need to take the Kerr effect into account, which gives rise to an additional contact interaction term in the equation for the photon BEC wave function. In total, the steady state of the condensate is, thus, described by
\begin{align}\label{eq:photon_ansatz}
\mu\psi = \left(-\frac{\hbar^2\nabla^2}{2m}+V+g_K|\psi|^2+\gamma \Delta T\right)\psi\, ,
\end{align}
where $m$ represents the effective photon mass and $V$ describes the external potential. The strength of the Kerr interaction is given by $g_K$ and the energy shift due to the temperature difference $\Delta T$ between the actual intracavity temperature and the room temperature is intermediated by the parameter $\gamma$, which is proportional to the thermo-optic coefficient $\beta$ from figure~\ref{Fig:interaction} \cite{Stein2019}. Furthermore, the photon BEC wave function is normalised according to $\int d^2x~|\psi|^2=N$.\\
The steady state of the temperature difference $\Delta T$, which is produced by the photon condensate due to non-perfect absorption-reemission cycles and which diffuses through the cavity, is described by the diffusion equation
\begin{align}\label{eq:temp_ansatz}
\Delta T = \tau\mathcal D\nabla^2\, \Delta T + \sigma \tau B|\psi|^2\, .
\end{align}
Here $\tau$ denotes the longitudinal relaxation time stemming from the diffusion along the optical axis, see reference~\cite{Stein2019} and the appendix therein, $\mathcal{D}$ stands for the diffusion coefficient of the temperature, and the heating of the dye solution is modelled via the heating rate $B$. Furthermore, the duty cycle $\sigma$ describes that the experiment operates with a pulsed pump laser, whereas our theoretical description works with a continuous pump for reaching the steady state. This modification is needed here, as the temperature necessitates several experimental cycles to achieve its steady state \cite{Stein2019}.

\subsection{Elimination of Temperature Difference}
As a first step, we eliminate the temperature difference as a degree of freedom from our description.
To this end, we formally solve the diffusion equation \eqref{eq:temp_ansatz} according to  
\begin{align}\label{eq:TGreen}
\Delta T(\vec x) = \sigma \tau B \int d^2x'~\mathcal{G}(\vec x -\vec x') |\psi(\vec x')|^2\,,
\end{align}
where we have introduced the Green's function $\mathcal{G}(\vec x)$. Its Fourier transform $\tilde{\mathcal G}(\vec k)$ reads
\begin{align}\label{eq:green}
\tilde{\mathcal G}(\vec k) = \frac{1}{\tau\mathcal{D}\vec k^2+1}\, ,
\end{align}
so we conclude for the real space
\begin{align}
\mathcal G(\vec x) = \int \frac{d^2k}{4\pi^2}~\frac{e^{i\vec k\cdot\vec x}}{\tau\mathcal D\vec k^2+1} \, .
\end{align}
In order to evaluate the integral we use the Schwinger parametrisation \cite{crit-prop}
\begin{align}
\int_0^\infty dt~e^{-at}=\frac{1}{a}
\end{align}
and have then
\begin{align}
	\mathcal G(\vec x) = \int_0^\infty dt~\int \frac{d^2k}{4\pi^2}~e^{-(1+\tau\mathcal D \vec k^2)t + i\vec k\cdot \vec x}\,.
\end{align}
As the integral over $\vec k$ represents now  a Gaussian, we can calculate it and find 
\begin{align}\label{eq:green_x}
\mathcal{G} (\vec x) = \int_0^\infty dt~\mathcal G(\vec x, t),
\end{align}
with the integrand 
\begin{align}\label{eq:green_t}
	\G(\vec x,t) = 	\frac{e^{-\vec x^2/(4\ldiff^2t) -t}}{4\pi\ldiff^2 t}\, .
\end{align}
Here $\ldiff = \sqrt{\tau\mathcal D}$ represents the diffusion length and the Schwinger parameter $t$ corresponds physically to the time in units of the longitudinal relaxation time $\tau$. We recognise expression \eqref{eq:green_t} to be the Green's function of the time-dependent diffusion equation. Whereas at initial time the Green's function \eqref{eq:green_t} reduces to the delta function, i.e.,
\begin{align}\label{eq:green_t-dirac}
	\G(\vec x,0) = 	\delta (\vec x)\, ,
\end{align}
summing \eqref{eq:green_t} over all times finally yields the steady-state Green's function \eqref{eq:green_x}.
Evaluating the remaining Schwinger integral in equation \eqref{eq:green_x} leads to a modified Bessel function of the second kind $K_0$ \cite[(3.471.9)]{Gradshteyn2007}: 
\begin{align} \label{eq:explicit}
	\G(\vec x) = \frac{||{\vec x}||}{4\pi\ldiff^2}\,K_0\left(\sqrt{\frac{||{\vec x}||}{\ldiff}}\,\,\right)\,.		
\end{align}
Whereas the initial Green's function \eqref{eq:green_t-dirac} has its maximum at the origin ${\vec x}={\vec 0}$, the steady-state Green's function \eqref{eq:explicit} is maximal at a circle, whose radius is given by $||\vec x||\sim\ldiff$. Although we have an explicit expression \eqref{eq:explicit} for the Green's function \eqref{eq:green_x}, the Schwinger integral representation \eqref{eq:green_t} turns out to be more advantageous for the following analytic calculation, such that we prefer to use it instead throughout the remainder of this paper. Taking this into account, equation \eqref{eq:TGreen} can be written as
\begin{align}\label{eq:temp_final}
\Delta T(\vec x) = \sigma \tau B \int_0^\infty dt~\int d^2x'~\mathcal{G}(\vec x -\vec x', t) |\psi(\vec x')|^2\,.
\end{align}
With this the steady-state profile of the temperature difference is given due to diffusion by the photon density.

\subsection{Photon Functional}
Using the formal solution of the diffusion equation \eqref{eq:temp_ansatz} in the form \eqref{eq:temp_final}, the photon BEC wave function equation \eqref{eq:photon_ansatz} goes over into
\begin{align}\label{eq:photon}
\mu\psi = \left[-\frac{\hbar^2\nabla^2}{2m}+V+g_K|\psi|^2+g_T \int_0^\infty dt~\int d^2x'~\mathcal{G}(\vec x -\vec x', t) |\psi(\vec x')|^2  \right]\psi\,.
\end{align}
Here the resulting thermo-optic interaction strength is defined as \cite{Stein2019}
\begin{align}
g_T = \sigma\gamma\tau B,
\end{align}
and is, thus, determined by various material properties of the dye solution. As a next step, we determine the energy functional corresponding to equation \eqref{eq:photon}, which turns out to consist of three parts: 
\begin{align}\label{eq:functional}
	E[\psi^*,\psi] =E_0[\psi^*,\psi] + E_K[\psi^*,\psi] + E_{T}[\psi^*,\psi]\,.
\end{align}
The first one describes both the kinetic and the potential energy of the photon BEC and reads
\begin{align}\label{eq:e0}
    E_0[\psi^*,\psi] = \int d^2x~\left[\frac{\hbar^2}{2m}|\nabla\psi|^2 + V|\psi|^2\right]\,,
\end{align}
whereas the second one, 
\begin{align}\label{eq:eg}
    E_K[\psi^*,\psi] = \frac{g_K}{2}\int d^2x~|\psi|^4,
\end{align}
represents the contact Kerr interaction. The last term comprises the thermo-optic effects via
\begin{align}\label{eq:ed}
    E_{T}[\psi^*,\psi] = \frac{g_T}{2}\int_0^\infty dt\int d^2x\int d^2x'~\mathcal{G}(\vec x -\vec x', t) |\psi(\vec x')|^2|\psi(\vec x)|^2\,.
\end{align}
In the following we aim at minimising the energy functional \eqref{eq:functional} for a harmonic confinement along the dimensional crossover within a variational approach, similar to our preceding work \cite{Stein2019}.

\section{Variational Approach}\label{sec:minimisation}
We express the harmonic potential in the form
\begin{align}\label{eq:V}
    V = \frac{m\Omega^2}{2}(x^2+\lambda^4y^2),
\end{align}
where the trap aspect ratio $\lambda = l_x/l_y$ determines the ratio of the oscillator length $l_i = \sqrt{\hbar/(m\Omega_i)}$ with $i=x,y$ in the respective dimensions and $\Omega=\Omega_x$ is the trapping frequency in $x$-direction.
As the photon condensate wave function is a Gaussian in the non-interacting case, it is reasonable to assume that this profile remains to be valid also in the mutual presence of both Kerr and thermo-optic interaction. Therefore, the variational ansatz for the phBEC ground-state wave function reads 
\begin{align}\label{eq:ansatz_ho_ground}
	\psi = \sqrt{\frac{\lambda N}{\alpha_x\alpha_y\pi l_x^2}} \exp{\left[-\frac{1}{2l_x^2}\left(\frac{x^2}{\alpha_x^2}+\lambda^2\frac{y^2}{\alpha_y^2}\right)\right]},
\end{align}
where we treat $\alpha_x, \alpha_y$ as the corresponding variational parameters. Note that due to this choice, these parameters are dimensionless and $\alpha_x=\alpha_y=1$ describes the non-interacting case. Inserting the ansatz \eqref{eq:ansatz_ho_ground} into the functional \eqref{eq:functional} yields the energy as a function of the two variational parameters and the ratio $\lamdiff= l_x/\ldiff$ of the oscillator length $l_x$ and the diffusion length $\ldiff$:
\begin{align}\label{eq:E_Var}
	\begin{split}
  E(\alpha_x,\alpha_y) = N\hbar\Omega &\Bigg[\frac{1}{4}\left(\frac{1}{\alpha_x^2}+\frac{\lambda^2}{\alpha_y^2}\right) + \frac{1}{4}\left(\alpha_x^2+\lambda^2\alpha_y^2 \right)+
    \frac{\tilde g_K \lambda N}{4\pi \alpha_x\alpha_y} \\
				      &+ \frac{\tilde g_T\lambda N}{4\pi\alpha_x\alpha_y}\int_0^\infty dt\,\frac{e^{-t}}{\sqrt{[1+2t/(\lamdiff^2\alpha_x^2)][1+2t\lambda^2/(\lamdiff^2\alpha_y^2)]}} \Bigg]\, .
	\end{split}
\end{align}
Note that we have defined here the dimensionless interaction strength $\tilde g_\bullet = mg_\bullet/\hbar^2$. Thus, by performing the derivative of the function \eqref{eq:E_Var} either with respect to $\alpha_x$ or with respect to $\alpha_y$ we can calculate the corresponding equations for the variational parameters and obtain
\begin{align}\label{eq:ax}
  \alpha_x = \frac{1}{\alpha_x^3} + \frac{\tilde g_K\lambda N}{2\pi \alpha_x^2\alpha_y} + \frac{\tilde g_T\lambda N}{2\pi\alpha_x^2\alpha_y} \int_0^\infty dt\,
  \frac{e^{-t}}{\sqrt{[1+2t/(\lamdiff^2 \alpha_x^2)]^3[1+2t\lambda^2/(\lamdiff^2\alpha_y^2)]}} 
\end{align}
for the $x$ direction and in the squeezed $y$ direction we have
\begin{align}\label{eq:ay}
  \lambda^2\alpha_y = \frac{\lambda^2}{\alpha_y^3} + \frac{\tilde g_K\lambda N}{2\pi\alpha_x\alpha_y^2} + \frac{\tilde g_T\lambda N}{2\pi\alpha_x\alpha_y^2} \int_0^\infty dt\,
  \frac{e^{-t}}{\sqrt{[1+2t/(\lamdiff^2 \alpha_x^2)][1+2t\lambda^2/(\lamdiff^2\alpha_y^2)]^3}}.
\end{align}

\subsection{General Solution}
\begin{figure}
	\centering
	\includegraphics[width=\linewidth]{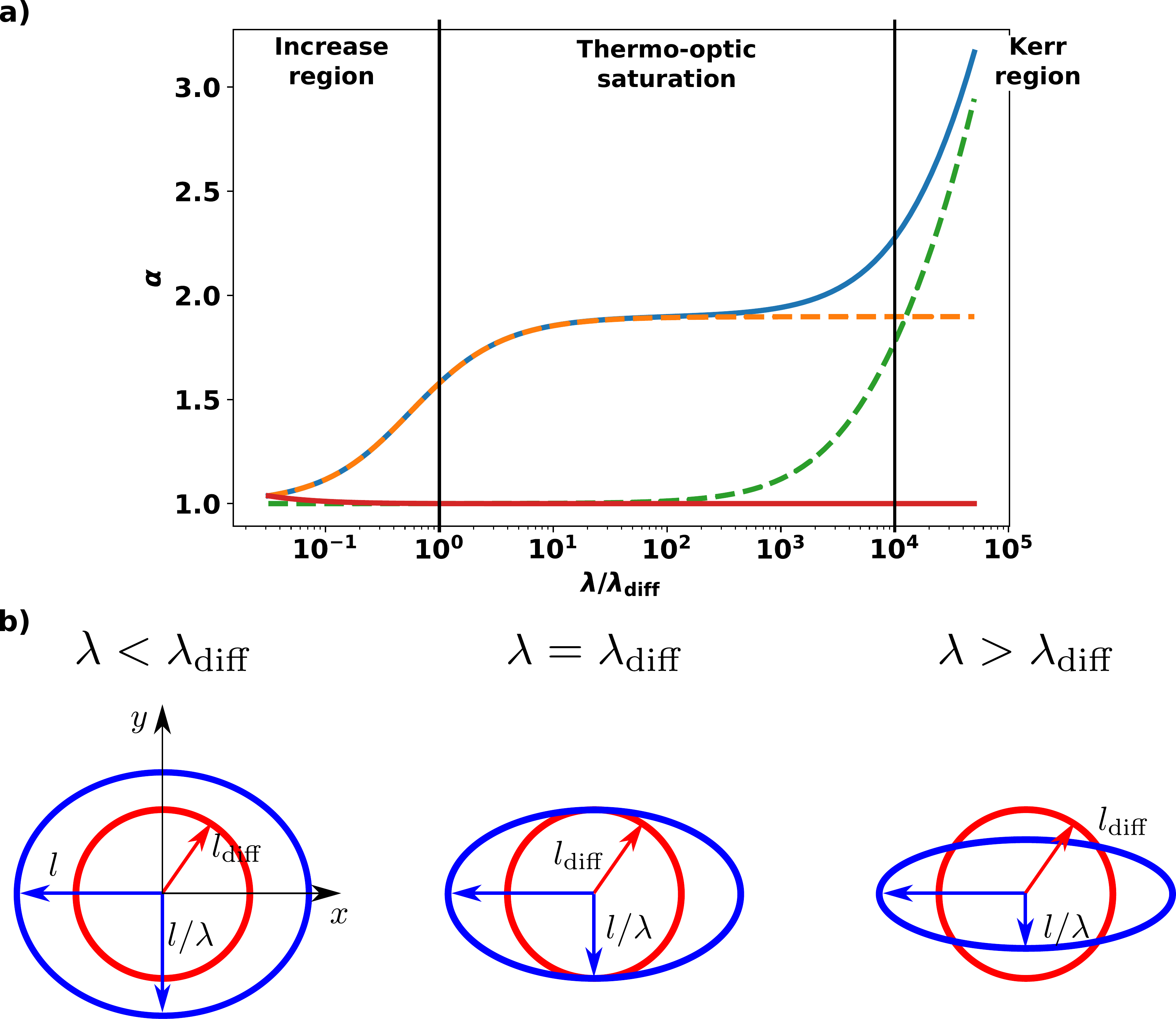}
	\caption{Numerical solution of equations \eqref{eq:ax} and \eqref{eq:ay} with experimental parameters $N=10^4$, $\tilde g_K=10^{-8}$, $\tilde g_T=10^{-4}$, and $\lamdiff=32$ for varying trap aspect ratio $\lambda$. \textbf{a)} Variational parameters $\alpha_x$ (blue) and $\alpha_y$ (red). The dashed lines take merely the thermo-optic (orange) and the Kerr (green) influence upon $\alpha_x$ into account. \textbf{b)} Schematic representation of length scales of the condensate $l$ and $l/\lambda$ in different directions in comparison with the diffusion length scale $\ldiff$ for different values of the trap aspect ratio $\lambda$. }
	\label{Fig:widths_general}
\end{figure}

At first, we discuss the general numerical solution of equations \eqref{eq:ax} and \eqref{eq:ay} as depicted in figure~\ref{Fig:widths_general} a). We see that, as the trap aspect ratio $\lambda$ increases, the variational parameter $\alpha_y$ approaches the value 1. This indicates that in this direction the broadening due to the interaction gets negligible, which means that the system behaves effectively one-dimensional. On the other hand, we observe a much more complex behaviour for the variational parameter $\alpha_x$, where we discern in total three regions. For small trap aspect ratios $\lambda$ the parameter $\alpha_x$ starts to grow, which is a characteristic sign of increasing interaction. Then, for intermediate $\lambda\sim\lamdiff$, we find that the variational parameter $\alpha_x$ saturates, which signals a saturation of the interaction. And finally, for large trap aspect ratio $\lambda$, the variational parameter $\alpha_x$ increases again. We can understand this behaviour in more detail by separating the different interaction mechanisms numerically. The green and the red dashed line show the width by only taking the thermo-optic interaction and the Kerr interaction into account, respectively. We note, that, indeed, the thermo-optic interaction is the dominant interaction effect for small $\lambda$ and saturates at $\lambda\sim\lamdiff$. The Kerr interaction, on the other hand, behaves differently. Its contribution for small trap aspect ratio $\lambda$ is negligible, but becomes stronger than the thermo-optic interaction at $\lambda\gtrsim \lambda_\text{Kerr}=\lamdiff\, g_T/g_K$. This threefold behaviour is schematically shown in figure~\ref{Fig:widths_general} b) by depicting the length scales of the condensate $l$ and $l/\lambda$ in different directions in comparison with the diffusion length scale $\ldiff$ for different values of the trap aspect ratio $\lambda$. Note that the particular role of the diffusion length scale $\ldiff$ can be traced back to the steady-state Green function \eqref{eq:explicit}, which is maximal at the circle with radius proportional to $\ldiff$. In the following we discuss these findings in more detail.

\subsection{Isotropic Case \label{ssec:isotrop}}
In the isotropic case we have $\lambda=1$ and $\alpha_x = \alpha_y =:\alpha$, so equations \eqref{eq:ax} and \eqref{eq:ay} reduce to the single equation
\begin{align}
    \alpha = \frac{1}{\alpha^3} + \frac{\tilde g_KN}{2\pi\alpha^3} + \frac{\tilde g_TN}{2\pi\alpha^3}\int_0^\infty dt\,\frac{e^{-t}}{(1+2t/\lamdiff^2\alpha^2)^2}\,.
\end{align}
For the parameters of the Bonn experiment \cite{Klaers2010, ApplPhysB} we estimate that $\lamdiff^2\sim 10^3\gg1$. Furthermore, the exponential in the integral leads to an effective cutoff of the integral for $t\sim 1$. Then the term $2t/\lamdiff^2\alpha^2$  in the denominator can be neglected, since it only contributes to the integral at times $t\sim\alpha^2\lamdiff^2/2\gg 1$. With this we can calculate approximately the integral and conclude that $\alpha$ is determined by the algebraic equation
\begin{align}\label{eq:iso}
    \alpha \approx \frac{1}{\alpha^3} + \frac{\tilde g_KN}{2\pi \alpha^3} + \frac{\tilde g_TN}{2\pi \alpha^3}\,.
\end{align}
Here the thermo-optic interaction behaves exactly as the Kerr interaction, as the influence of the diffusion dropped out. Furthermore, we can solve equation \eqref{eq:iso} for the variational parameter and obtain
\begin{align}
  \alpha = \sqrt[4]{1+\frac{(\tilde g_K+\tilde g_T) N}{2\pi}}\,,
\end{align}
which we already obtained in the former work \cite{Stein2019}.

\subsection{Quasi 1D Case \label{ssec:q1D}}
Now we deal with the opposite situation, where the system is quasi-one-dimensional and determine at first for which trap aspect ratios $\lambda$ this regime starts. To this end we read off from figure~\ref{Fig:widths_general} a) that already for $\lambda  \ll \lamdiff$ we have $\alpha_y \approx 1$, which is a sign that in the squeezed direction the influence of the interaction is negligible. In this case the sum of kinetic and potential energy $E_y \approx N \hbar \Omega\lambda^2 /2$ stored in the squeezed spatial degree of freedom is proportional to the square of the trap aspect ratio, whereas the interaction energy $E_\text{int} = E_K + E_T$ increases linearly with $\lambda$ according to expression \eqref{eq:E_Var}. Indeed, in this regime the contribution of the Kerr interaction can be neglected in comparison to the thermo-optic interaction according to figure~\ref{Fig:widths_general} a). For the remaining integral we can apply the same approximation as in section \ref{ssec:isotrop}, resulting finally in $E_\text{int} \approx \hbar\Omega\tilde g_T\lambda N/(4\pi\alpha_x)$. A further inspection of figure~\ref{Fig:widths_general} reveals that we can roughly approximate $\alpha_x\approx1$ in this regime as well. Thus, as the quasi-1D region amounts to the inequality $E_y\gg E_\text{int}$, we obtain the criterion
\begin{align}\label{eq:1D-criterion}
	\lambda\gg\lambda_\text{1D}=\frac{\tilde g_TN}{2\pi}\,.
\end{align}
As current photon BEC experiments are characterized by $\tilde g_T=10^{-4}$ and a maximal photon number $N=10^5$, the 1D criterion \eqref{eq:1D-criterion} is basically fulfilled slightly above the 2D case $\lambda=1$.\\
We proceed now to larger values of the trap aspect ratio $\lambda$, where we can still assume $\alpha_y\approx1$, according to figure~\ref{Fig:widths_general} a). We can now determine $\alpha_x$ self-consistently from equation \eqref{eq:ax}, yielding
\begin{align}
  \alpha_x\approx \frac{1}{\alpha_x^3} + \frac{\tilde g_K\lambda N}{2\pi\alpha_x^2} + \frac{\tilde g_T\lambda N}{2\pi \alpha_x^2}\int_0^\infty dt\,
  \frac{e^{-t}}{\sqrt{(1+2t/\lamdiff^2\alpha_x^2)^3(1 + 2t\lambda^2/\lamdiff^2)}}\,.
\end{align} 
This integral is simplified along similar lines as in section \ref{ssec:isotrop}, and it reduces to
\begin{align}
    \alpha_x^4\approx 1 + \frac{\tilde g_\text{1D}(\lambda) N}{\sqrt{2\pi}}\,\alpha_x\,,
\end{align}
where we have introduced the effective 1D interaction strength inspired by a comparison with equation \eqref{eq:realE1D}:
\begin{align}\label{eq:g1D}
	\tilde g_\text{1D}(\lambda) = \frac{1}{\sqrt{2\pi}}\left[\tilde g_K\lambda + \tilde g_T \lamdiff\sqrt{\frac{\pi}{2}}e^{\lamdiff^2/(2\lambda^2)} \erfc{\frac{\lamdiff}{\sqrt{2}\lambda}} \right]\,.
\end{align}
We note that the contribution of the thermo-optic interaction is determined by the ratio $\lambda/\lamdiff = \lambda \ldiff/l_x$, i.e., the ratio of the diffusion length $\ldiff$ and the oscillator length in the squeezed $y$ direction $l_x/\lambda$. 
\begin{figure}
	\centering
	\includegraphics[width=.7\linewidth]{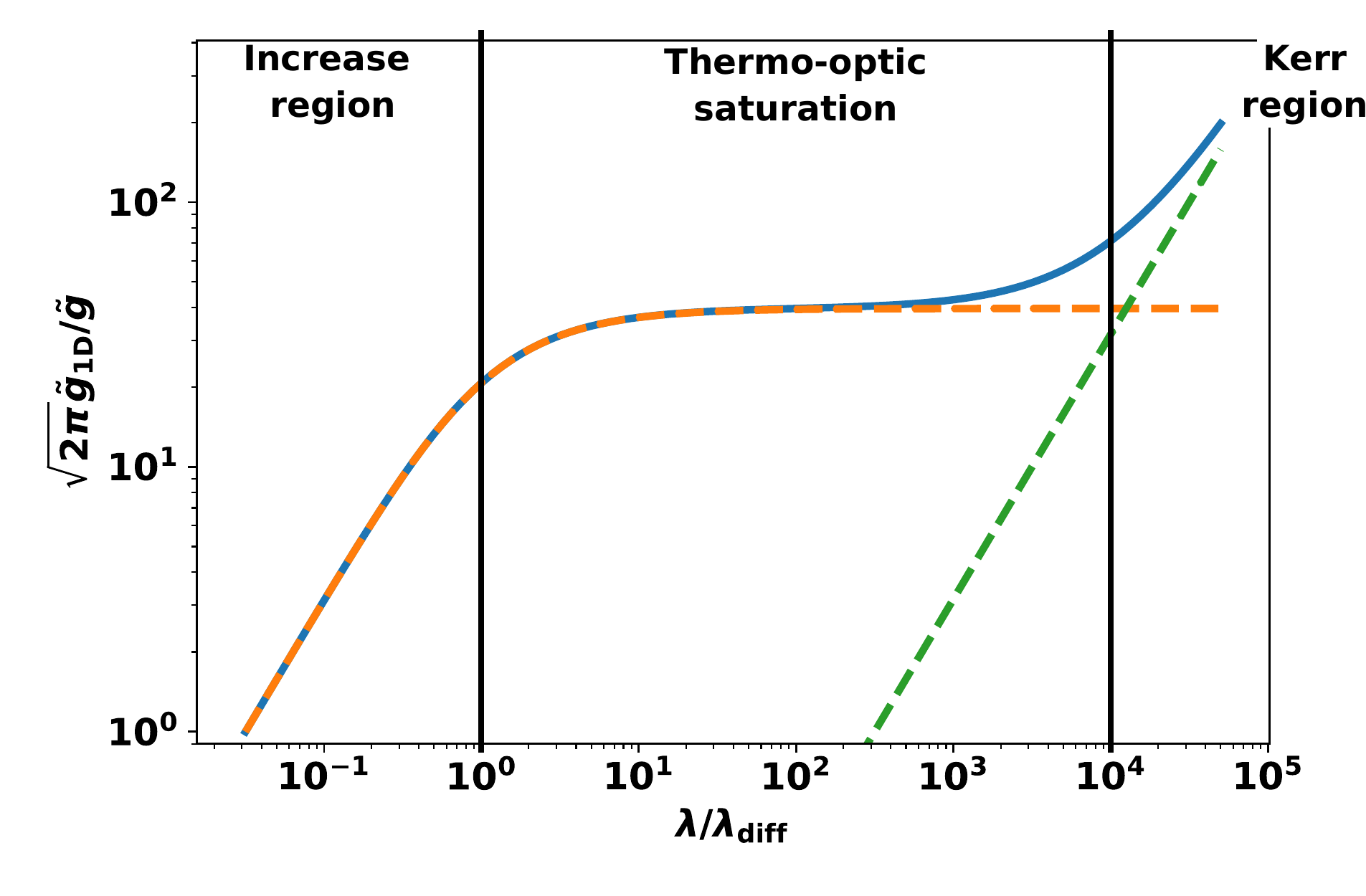}
	\caption{Effective one-dimensional interaction strength $\tilde g_\text{1D}(\lambda)$ from definition \eqref{eq:g1D} normalised to the isotropic 2D interaction constant $\tilde g = \tilde g_K+\tilde g_T$ in blue for the experimental parameters $\tilde g_T = 10^{-4}$, $\tilde g_K=10^{-8}$, and $\lamdiff=32$. The dashed orange line shows the thermo-optic contribution, whereas the dashed green line depicts the contribution of the Kerr interaction.}
	\label{Fig:g1D}
\end{figure}
Figure \ref{Fig:g1D} depicts the total effective 1D interaction strength $\tilde g_{1\text{D}}(\lambda)$ from equation \eqref{eq:g1D} as a function of $\lambda$. Also here we note the aforementioned three different regions of the crossover. For small trap aspect ratio $\lambda$ the thermo-optic interaction, which gives here the leading contribution, increases and then indeed, as stated above, saturates. But for $\lambda>\lambda_\text{Kerr}=\lamdiff\, \tilde g_T/\tilde g_K$  the Kerr interaction takes over and the total interaction grows again.\\
Let us now discuss these findings in more detail. For small trap aspect ratio, i.e., $\lambda  \ll \lamdiff$, we can approximate equation \eqref{eq:g1D} by
\begin{align}
	\tilde g_{\text{1D},0}(\lambda) \approx \frac{1}{\sqrt{2\pi}}\left(\tilde g_K + \tilde g_T\right)\lambda\,,		
\end{align}
and in this case the thermo-optic interaction behaves like the Kerr interaction showing a linear increase in $\lambda$. The reason for this is that here the diffusion length is negligible compared to the condensate width in both $x$ and $y$ direction. Thus, the heat produced by the condensate only diffuses within a region where the condensate wave function does not vary, such that the thermo-optic interaction behaves approximately as a local contact interaction. On the other hand, once we have entered deeply the quasi-1D regime, i.e., $\lambda  \gg \lamdiff$, the effective 1D interaction strength \eqref{eq:g1D} is given by  
\begin{align}\label{eq:g_inf}
	\tilde g_{\text{1D},\infty}(\lambda) \approx \frac{1}{\sqrt{2\pi}}\left(\tilde g_K\lambda + \tilde g_{T, \infty}\right)\,.
\end{align}
Thus, in this limit the thermo-optic part of the interaction strength no longer depends on the trap aspect ratio $\lambda$ and saturates at the value
\begin{align} \label{eq:g_max}
	\tilde g_{T,\infty} = \tilde g_T\lamdiff\,,
\end{align}
which is fixed by the geometry of the experiment and by the used solvent. This is due to the fact that here the width of the condensate in the squeezed direction $l_x/\lambda$ is much smaller than the diffusion length $\ldiff$ and, thus, the heat being produced by the condensate diffuses through the dye medium to regions where no condensate exists, cf.~figure~\ref{Fig:widths_general}. This heat, therefore, cannot contribute to the interaction, such that the thermo-optic interaction saturates. For the Kerr contribution, however, the situation does not change and the total interaction strength still shows according to equation \eqref{eq:g_inf}) a linear dependency in the trap aspect ratio $\lambda$.\\
Note that it is currently reasonably to expect achieving experimentally a trap anisotropy of at most  $\lambda \sim 10^2\lamdiff$ \cite{PrivFrank}. From figure~\ref{Fig:interaction} we read off that in this case the Kerr interaction is still negligible and that the total effective 1D interaction is due to the thermo-optic effect. Thus, the maximally achievable effective 1D interaction strength $\tilde g_{\text{1D},\infty}^\text{exp}$ reads, with the help of expressions \eqref{eq:g_inf} and \eqref{eq:g_max}
\begin{align}
	\tilde g_{\text{1D},\infty}^\text{exp} = \frac{\lamdiff}{\sqrt{2\pi}} \tilde g_T.
\end{align}
Therefore, we can, indeed, expect an increase of the effective photon-photon interaction strength via a dimensional crossover. Taking into account that $\lamdiff\sim{32}$, the expected increase of the interaction strength amounts to more than one order of magnitude.

\subsection{Energy}
\begin{figure}
	\centering
	\includegraphics[width=.7\linewidth]{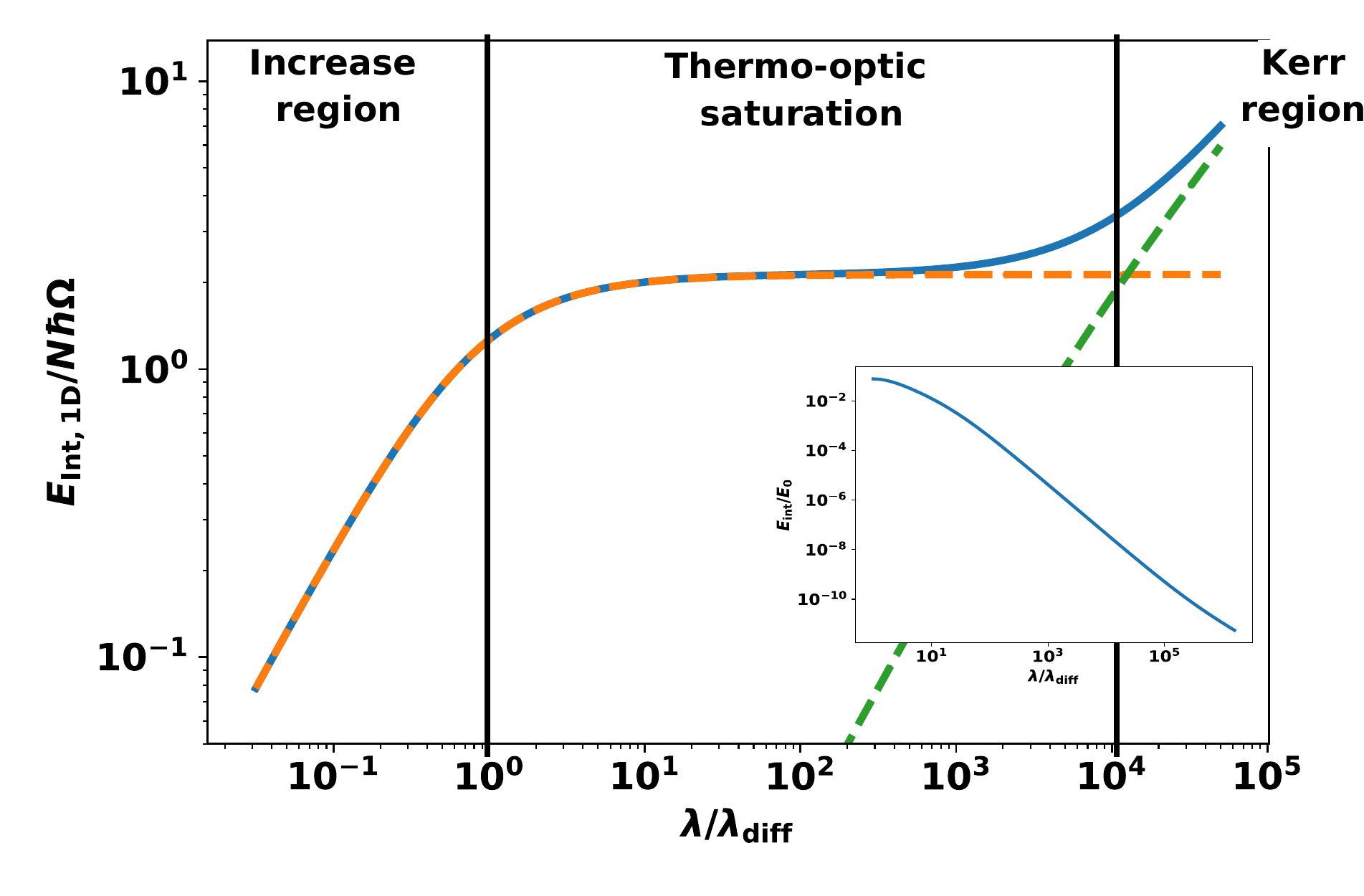}
	\caption{Interaction energy $E_\text{int, 1D}$ given by equation \eqref{eq:Eint} for the experimental parameters $N=10^4$, $\tilde g_K=10^{-8}$, $\tilde g_T=10^{-4}$, and $\lamdiff=32$. The orange dashed line shows the contribution of the thermo-optic effect, whereas the green dashed line indicates the contribution of the Kerr effect. The inset shows the interaction energy relative to the non-interacting energy $E_0$ from \eqref{eq:E0}.}
	\label{Fig:E1D}
\end{figure}

From analysing the behaviour of the variational parameters, which are basically the widths of the phBEC wave function, at the dimensional crossover it is obvious that the effective photon-photon interaction strength can be measured quite directly. However, we emphasise that this measurement relies on evaluating real space images of the light leaking out the cavity. More precise results are expected from spectroscopic measurements of this light, which directly reveals the phBEC energy. Thus, we discuss now the resulting energy of the condensate in more detail. In the effective 1D case by using $\alpha_y\approx1$ and the definition \eqref{eq:g1D} of the effective 1D interaction strength from the energy function \eqref{eq:E_Var}, we find for the energy 
\begin{align}
	E_\text{1D} \approx \frac{N\hbar\Omega}{2}\left[\frac{1}{2}\left(\frac{1}{\alpha_x^2}+\alpha_x^2\right) + \lambda^2 + \frac{\tilde g_\text{1D}(\lambda)N}{\sqrt{2\pi}\alpha_x}\right]\,.	
\end{align}
This formally coincides with \eqref{eq:realE1D} from \ref{App:1DGPE} apart from the $\lambda^2$ dependency, which represents the shift of the ground state due to the energy of the squeezed direction. Introducing the non-interacting energy
\begin{align}\label{eq:E0}
	E_0 = \frac{N\hbar\Omega}{2}\left(1+\lambda^2\right),	
\end{align}
we can define the interaction contribution to the energy by
\begin{align}\label{eq:Eint}
	E_\text{int, 1D} = E_\text{1D} - E_0\,,	
\end{align}
which is plotted in figure~\ref{Fig:E1D} as a function of the trap aspect ratio $\lambda$.
Again, we find the same threefold behaviour we have already observed for the widths and the interaction strength, which stems from a saturation of the thermo-optic interaction for intermediate $\lambda$. Moreover, we see from the inset in figure~\ref{Fig:E1D} that the interaction energy $E_\text{int, 1D}$ is quite small compared to the unperturbed energy (\ref{eq:E0}), so that our variational approach is a good approximation of the true ground state.\\
Finally, we remark that our findings can be measured by utilising the fact that the thermo-optic interaction builds up steadily during the experimental run. At the beginning of the experiment the dye-filled solution in the cavity does not have any temperature difference with respect to the environment, so the thermo-optic interaction does not yet occur, whereas the instantaneous Kerr interaction is already fully present. As a single experiment lasts only about \SI{500}{\nano\second}, the temperature difference saturates only after several pump pulses, such that then the thermo-optic interaction is in its steady state and yields its full contribution. Consequently, the resulting strength of the thermo-optic interaction can be measured by determining the energy of the condensate at the beginning of the experiment and by comparing it with the energy at the end. In principle this would involve subtracting the Kerr contribution from the interaction energy \eqref{eq:Eint},
\begin{align}\label{eq:Eth}
	E_\text{th, 1D} = E_\text{int, 1D}-\frac{\hbar\Omega\tilde g_K\lambda N^2}{2\sqrt{2\pi}\alpha_x}\,.
\end{align}
However, as already mentioned above, in the experiment only the thermo-optic saturation region is expected to be accessible. According to  figure~\ref{Fig:E1D} the energy contribution due to the Kerr effect is negligible in the whole experimental regime. According to equation \eqref{eq:Eth} the total interaction energy \eqref{eq:Eint} coincides with the thermo-optic energy contribution. Therefore, one can directly use expression \eqref{eq:Eint} to determine the strength of the effective photon-photon interaction. We remark that for sufficiently small particle number $N$ one can enter the regime where it is valid to determine the variational parameters perturbatively in first order with respect to the smallness parameter $\tilde g_\text{1D}(\lambda)N/\sqrt{2\pi}$. In this case the interaction energy \eqref{eq:Eint} is directly given by 
\begin{align}
	E_\text{int, 1D}\approx	\frac{\hbar\Omega\tilde g_\text{1D}(\lambda) N^2}{2\sqrt{2\pi}}
\end{align}
allowing to directly determine the effective interaction strength from the measured value of $E_\text{int, 1D}$. 

\section{Summary}
In this paper we have shown how the ground state of a phBEC changes during the dimensional crossover from 2D to 1D. Our main focus in this investigation was the behaviour the effective photon-photon interaction strength in the crossover in order to make effects like superfluidity accessible in experiments. We have found that the effective photon-photon interaction strength increases through the crossover. However, we have shown that the thermo-optic interaction can only be increased up to a factor $\lamdiff/\sqrt{2\pi}$, cf.~section \ref{ssec:q1D}. The deeper physical reason behind this finding is that for large enough trap aspect ratio a large amount of energy is carried away by the heat diffusion from the region occupied by the condensate and cannot contribute to the interaction anymore. Contrarily to that, the Kerr interaction increases linearly with the trap aspect ratio such that for a large trap anisotropy, which is presumably not achievable in current experiments, the Kerr interaction gives the leading interaction effect. Therefore, we have shown that the effective photon-photon interaction may be increased by more than an order of magnitude compared to the currently available experiments in 2D.

\ack
We thank Antun Bala\v{z}, Georg von Freymann, Milan Radonji\'{c}, Julian Schulz, Kirankumar Karkihalli Umesh and Frank Vewinger for insightful discussions. E. S. and A. P. acknowledge financial support
by the Deutsche Forschungsgemeinschaft (DFG, German Research Foundation) via the Collaborative Research Center SFB/TR185 (Project No. 277625399).

\section*{References}
\bibliographystyle{unsrt}
\bibliography{refs}

\appendix

\section{1D Gross-Pitaevskii Equation \label{App:1DGPE}}
In order to compare the results from the dimensional crossover to the exact 1D scenario, we review in this section the steady state of a one-dimensional Gross-Pitaevskii equation with harmonic trapping potential. Thus, we have to solve
\begin{align}
	\mu \psi = \left(-\frac{\hbar^2\nabla^2}{2m}+\frac{m\Omega^2}{2}x^2+g_\text{1D}|\psi|^2\right)\psi
\end{align}
with the corresponding functional
\begin{align}\label{eq:func_ho_1d}
    E_\text{1D}[\psi,\psi^*] = \frac{\hbar^2}{2m}\int dx~\left|\partial_x\psi\right|^2 + \frac{m\Omega^2}{2}\int dx~x^2|\psi|^2 + \frac{g_\text{1D}}{2}\int dx~|\psi|^4\,.
\end{align}
In order to obtain an approximate solution, we use a Gaussian ansatz function
\begin{align}\label{eq:ansatz_ho_1d}
	\psi = \sqrt{\frac{N}{\sqrt{\pi}l\alpha}} e^{-x^2/(2l^2\alpha^2)}\,,
\end{align}
where  $l=\sqrt{\hbar/(m\Omega)}$ stands for the oscillator length and $\alpha$ represents the dimensionless variational parameter. Inserting ansatz \eqref{eq:ansatz_ho_1d} in the energy functional \eqref{eq:func_ho_1d} yields for the energy 
\begin{align}\label{eq:realE1D}
	E_\text{1D} = \frac{\hbar\Omega}{2}\left( \frac{1}{2\alpha^2} + \frac{\alpha^2}{2} + \frac{\tilde g_\text{1D}N}{\sqrt{2\pi}\alpha} \right)\, ,
\end{align} 
where we define the dimensionless 1D interaction strength \cite[sec.~15.3.2]{Pethick}
\begin{align}
\tilde{g}_\text{1D} = \frac{g_\text{1D}ml}{\hbar^2}\,.
\end{align}
Extremising \eqref{eq:realE1D} with respect to the dimensionless width $\alpha$, we obtain the algebraic equation
\begin{align}\label{eq:real1D}
    \alpha^4 = 1 + \frac{N\tilde{g}_\text{1D}}{\sqrt{2\pi}}\alpha\,.
\end{align}

\end{document}